# Earth-like Habitats in Planetary Systems

[1]Fritz J., [2]Bitsch B., [3]Kührt E., [2]Morbidelli A., [3]Tornow C., [1]Wünnemann K., [1]Fernandes V. A., [3]Grenfell, J. L., [3]Rauer H., [3]Wagner, R., [4]Werner S. C.

1 Museum für Naturkunde Berlin Invalidenstraße 43, 10115 Berlin Germany

2 Observatoire de la Côte d'Azur, Nice, France

3 Deutsches Zentrum für Luft und Raumfahrt, Berlin, Germany

4 Centre for Earth Evolution and Dynamics (CEED), University of Oslo, Norway

Corresponding author: Jörg Fritz - E-mail: joerg.fritz@mfn-berlin.de

**Abstract:** Understanding the concept of habitability is clearly related to an evolutionary knowledge of the particular planet-in-question. However, additional indications so-called "systemic aspects" of the planetary system as a whole governs a particular planet's claim on habitability. In this paper we focus on such systemic aspects and discuss their relevance to the formation of an "Earth-like" habitable planet. This contribution summarizes our results obtained by lunar sample work and numerical models within the framework of the Research Alliance "Planetary Evolution and Life". We consider various scenarios which simulate the dynamical evolution of the Solar System and discuss the consequences for the likelihood of forming an Earth-like world orbiting another star. Our model approach is constrained by observations of the modern Solar System and the knowledge of its history. Results suggest that on the one hand the long-term presence of terrestrial planets is jeopardized due to gravitational interactions if giant planets are present. On the other hand the habitability of inner rocky planets may be supported in those planetary systems hosting giant planets.

Gravitational interactions within a complex multiple-body structure including giant planets may supply terrestrial planets with materials which formed in the colder region of the proto-planetary disk. During these processes, water, the prime requisite for habitability, is delivered to the inner system. This may occur either during the main accretion phase of terrestrial planets or via impacts during a post-accretion bombardment. Results for both processes are summarized and discussed with reference to the lunar crater record.

Starting from a scenario involving migration of the giant planets this contribution discusses the delivery of water to Earth, the modification of atmospheres by impacts in a planetary system context and the likelihood of the existence of extrasolar Earth-like habitable worlds.

# 1. Introduction

The large variety of possible outcomes associated with planetary formation in an N-body system may result in a large number of processes and final states, one of which is habitable planets as testified by the existence of Earth. In a famous quote Aristotle (384-322 BC) stated that "*The whole is more than the sum of its parts*". Whether or not a planet becomes habitable depends on many factors; e.g., its place and time of formation and the nature of its central star(s), on the formation of the other planets in its neighbourhood and the interactions within the entire system and its surroundings. Some important processes in proto-planetary and planetary systems which can affect the resulting habitability are:

- Gravitational interactions which govern the collapse of a molecular cloud, the interaction of planets with the disk and N-body interactions in the ensemble of planets and planetesimals,

- Hydrodynamic interactions within the proto-planetary gas disk which include processes such as excitation of density waves and turbulence or the drag exerted by the gas onto small particles.

- Electromagnetic interactions which distribute energy via radiation of galactic and stellar origin within the solar nebula and cause erosion of solar nebula gas or planet atmospheres. In addition, radiation from the central star(s) delivers energy towards the planets and is responsible for the Yarkovsky, Yorp and Poynting-Robertson effects that aid in shaping the system.

- Collisional interactions between either charged particles (electrons or protons) within the solar wind and those objects in orbit around the central star(s), or low velocity collision and high velocity impacts between dust to planet sized objects which deliver energy and mass and/or sometimes even life (panspermia),

A large number of more-or-less random processes can "sculpt" complex planetary systems, during their evolution from a pre-planetary nebula up to the geological evolution of the individual planets. Thus, the Solar System, which hosts a modestly wet Earth-Moon system situated well within the "habitable zone" which extends around a medium sized star i.e., the Sun, has resulted from a chain of events which obviously occurred at just the right time, in the right place and with the right "intensity". However, such a serendipitous chain of events may not be characteristic of other planetary systems. The question naturally arises as to whether the system conditions which led to the formation of the Solar System are rare occurrences - this issue has of course immediate repercussions for the frequency of Earth-like planets in the Universe.

Consequently, this paper addresses the following questions:

- How does the planetary system shape the habitability of a single planet?

- What is the role of giant planets, impacts and water delivery in the formation of habitable planets?

- What does the lunar crater record tell us about the fossil abundance of Near Earth Objects (NEO) through time, and by implication about the orbital history of the Solar System?

- What can we learn in this context from extra-solar systems?

The paper starts with introducing different facets of habitability and the processes involved in transforming molecular clouds into planetary systems (Section 2). Section 3 presents models that have been developed during the course of the HGF Research Alliance (hereafter "the Alliance") to describe the formation and subsequent evolution of terrestrial planets. The goal was to identify those processes which "correctly came together" to make Earth a habitable planet. This exercise sets the scene for the discussion in Section 4 regarding the processes and timing of events related to the delivery of water to Earth, the role of impacts in shaping the planetary atmospheres and the likelihood for the existence of Earth-like worlds in extra-solar systems.

## 2. System aspects of habitability

### 2.1. On habitability and Earth-like habitats

Habitability is the ability to support life. The habitability of environments for "life as we know it" or "alien life forms" can either be viewed as an either-or quantity or rated with different scales (Frank et al., 2001; Lammer et al. 2009). Habitable environments in the universe evolve from and are influenced by their cosmic surroundings. The elements and isotopes that make up Earth and its inhabitants are generated during nucleosynthetic processes (Wasserburg et al., 2006) either through the galaxy (Diehl et al., 2006) or localized in star forming regions (Vasileiadis et al., 2009). This paper does not address cosmic limitations of habitable worlds (e.g., Gonzales, 2005; Gowanlock et al., 2011) but focuses on the formation of planets within the habitable zone (HZ) around a star. The Habitable Zone (HZ) is commonly defined as the distance of the habitable body to the central star (or multi star systems) where the stellar radiation allows for the presence of liquid water on the planet's surface (Whewell, 1853; Wallace, 1903; Kasting et al., 1993; Kopparapu et al. 2013). The Continuously Habitable Zone (CHZ) is the region within the HZ which remains habitable despite the change in stellar luminosity with time (see e.g., Kasting et al., 1993). In this paper we focus only on classical habitability (but see also e.g. Lammer et al., 2009 and references therein).

The concept of habitability refers to the environmental conditions that allow for the sustainability of life, but it does not require the actual presence of life. Nevertheless it is problematic to define a universal set of physical and chemical conditions for habitability, e.g., because different species have different needs and tolerances. This paper will not discuss habitability from the view point of theoretical biology. Instead it considers "habitable" only those environments that are favourable for "life as we know it", namely carbon based life requiring at least some liquid water and nutrients.

Besides Earth or Earth-like extra-solar planets, habitable conditions were thought to be present in more or less exotic environments like for example: 1) in restricted regions of the Venusian atmosphere, 2) in micro-habitats a few km below the fractured martian surface, where liquid water is likely to be present, 3) in a liquid water mantle below the crust of icy moons orbiting the outer giant planets. This contribution focuses on the systemic aspects regarding formation of a roughly Earth sized rocky and modestly wet planet in the HZ of a long lived star, such as the Sun. In other words a planet such as the Earth, that allows for the presence of habitable environments for "life as we know it" across the atmosphere/hydrosphere crust interfaces over long time scales (Class I planets; Lammer et al. 2009).

A planet with roughly Earth's size, composition and orbital position presents a paradox for planet formation theory. For the Sun it is expected that in the early phases of the proto-planetary disk the temperature at the location which eventually becomes the HZ was initially too high for the condensation of water (e.g. Morbidelli et al., 2000). Thus, instead of being formed exclusively from material from within the HZ, an Earth-like habitable planet needs either to accrete additional material from outside this hot region, or to move from outside into the HZ. This paradox illustrates that the formation of Earth-like habitable environments in a planetary system are not simply determined by the molecular "ingredients" the planet is initially composed of but also by the structure and evolution of the entire system.

## 2.2 Roots of habitability in molecular clouds and star formation

The general requirements for the evolution of "life as we know it" are water, carbon and nitrogen bearing molecules, and energy. Although they are rare in the interstellar medium all these components occur in the relatively denser molecular clouds. Interstellar photochemistry of the icy mantles on the dust grains enables the formation of complex organic substances, such as the amino acid precursor's acetonitrile and amino-acetonitril (Bernstein et al., 2002; Munoz Caro et al., 2002; Danger et al., 2011). However, for the development of complex biomolecules the surfaces of these interstellar dust grains may be too small. Larger objects can form as a result of the gravitational collapse of a molecular cloud.

Overall there is a high probability that the molecular cloud surrounding the parental core from which the Solar System formed was a stellar cluster that in addition gave rise to other stellar systems (Dukes and Krumholz, 2012), e.g., similar to the reflection nebula NGC 1333 in the Pereus molecular cloud (Walawender et al., 2008) or the M67 open cluster (Önehag et al., 2011). Studies of the collapse of molecular clouds and the evolution of a sub-cluster within have indicated that the inflowing gas associated with proto-stellar disks which form in a cluster environment accretes less angular momentum compared to stars which form in an isolated mode (Hayfield et al., 2011). This is partly caused by the competitive accretion in a cluster forming cloud. Consequently, the disks of gas and dust which surround those proto-stars in stellar clusters tend to be globally stable while isolated systems show fragmentation of the circumstellar disk. Nevertheless, a globally stable disk can still lead to giant planet formation due to local gravitational instabilities. Indeed, massive disks can become gravitationally unstable in their outer parts, which can trigger giant planet formation through gravitational collapse (Boss, 1997). Thus, if one assumes that the existence of giant planets in the outer region is a factor which favours habitability (see Section 4.3), then a clustered mode of star formation seems to be advantageous.

The central star(s) that forms during the collapse of a molecular cloud governs the habitability of the orbiting planets in several ways. The metallicity, stellar mass, age, luminosity, cosmic ray output and the star's position in the galaxy are all potentially relevant factors that affect collapse. The Sun is in many ways not a typical G-star (Gustafsson, 2008) e.g.it is a single star with an unusually constant luminosity output and a relatively high mass. About ~15% of Sun-like stars have a chemical composition similar to the Sun (Melendez et al., 2009; Ramırez et al., 2009). The stellar mass and luminosity may strongly affect the position and the extent of the HZ around a single star (Kasting et al., 1993) and around multi-stellar systems (Kane and Hinkel, 2013).

## 2.3 Structure of extra-solar systems

The advent of extra-solar planet detection allows one to investigate whether the Solar System – with its habitable Earth - is a common or a rare outcome of planetary system formation. The gas giants in the Solar System reside outside 5 AU (astronomical units), and the inner planets are of low-mass terrestrial type inside 1.7 AU. However, even the first detections of extra-solar planets in the 1990s (Mayor and Queloz, 1995) challenged the idea that the Solar System is a faithful proto-type of planetary systems in general. Reviews on the observational techniques to identify extra-solar planets can be found in Santos (2008), Dominik (2012), Mayor and Queloz (2012) and Rauer et al. (2013). A large diversity of planets and planetary systems emerged with the rapidly increasing number of detected extra-solar planets. These observed extra-solar planetary systems host several new classes of planets that are not found in the Solar

System. 'Hot-Jupiters' represent giant gas planets orbiting close to their host stars with orbital periods of a few days only. Gaseous 'hot-Neptunes' and rocky hot 'super-Earths' (having planetary masses in the range 1 $M_{Earth}$ < m ≤ 10 $M_{Earth}$) have been detected in recent years as detection sensitivities increased. These hot planets orbit their stellar hosts within the orbit of Mercury when compared to the Solar System. Included amongst these close-in planets are some of the smallest and best characterized rocky planets detected so far, such as GJ 1214b (Anglada-Escudé et al., 2013), CoRoT-7b (Hatzes et al., 2011) and Kepler-10b (Batalha et al., 2011); Kepler 37b (Barclay et al., 2013), and the two planets around Kepler 62 (Borucki et al., 2013). Fig. 1 shows planetary masses versus semi-major axes for all confirmed extra-solar planets (status 2013).

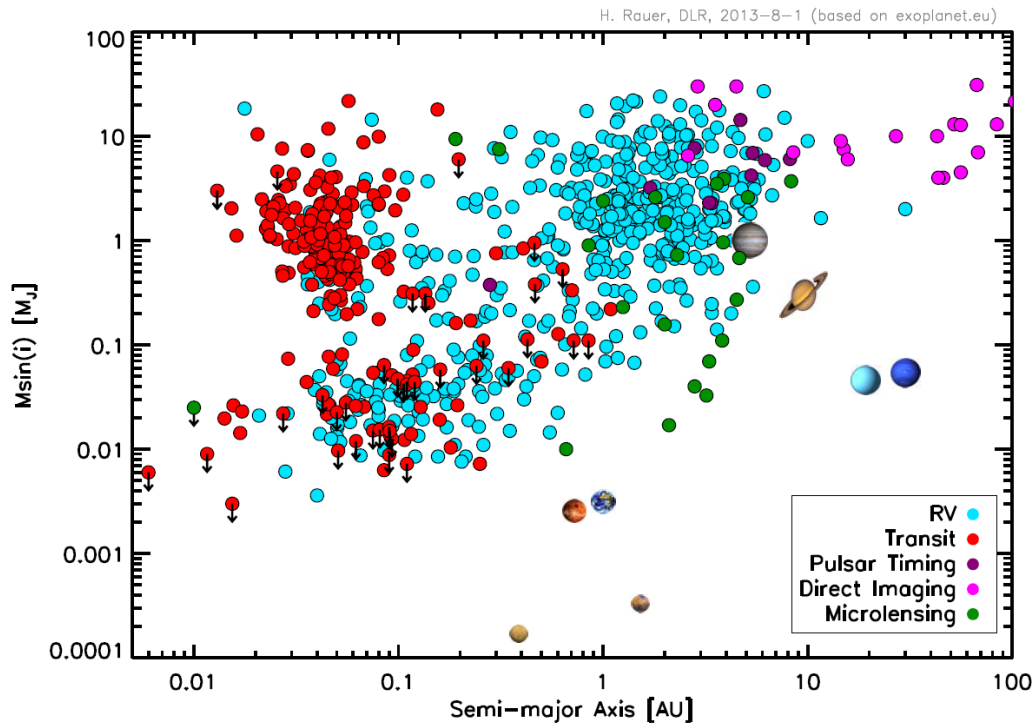

*Figure 1 Confirmed planet detections (March 2013). No planet candidates are included here. The red dots with downward arrows indicate those planets detected by the Kepler mission, where only upper limits of their mass are available (Rauer et al., 2013).*

Most of the detected rocky planets are on orbits with distances closer than the HZ of their central star (see Rauer et al., 2013 for an overview). However, some detections of rocky planets at intermediate distances have already been achieved, e.g., the planets in the GJ 876 (Rivera et al., 2005), GJ 667C (Anglada-Escudé et al., 2012) and Gliese 581 (Selsis et al., 2007) systems. Kepler-22b is orbiting within the HZ of a Sun-like star (Borucki et al., 2012) and the Kepler-62 system hosts two planets orbiting within the HZ of their central star (Borucki et al., 2013). In total about a dozen potentially rocky planets in the HZ are identified, although for most of them it is not determined whether as they are either super-Earths (rocky) or mini-Neptunes (gas planets). Identification as a rocky terrestrial planet requires measurements of both, radii and mass. Unfortunately, it is currently not possible to determine both mass and size for the planets mentioned above, because either these planets do not transit (so no radius measurement is available) or they orbit stars which are too faint; i.e., the planetary mass cannot be determined using radial velocity measurements. To guarantee a complete detection and characterization of rocky planets within the HZ of their host star(s) requires data from future missions, such as CHEOPS (ESA), TESS (NASA) and PLATO 2.0 (ESA).

Ground-based radial-velocity surveys and the Kepler mission have resulted in the publication of about 3000 planet candidates. This suggests that small planets could be more numerous than gas giants (e.g., Mayor et al. 2011; Howard 2013). It is estimated that 18% of stars harbour Earth size planets in orbital periods of up to 85 days, while large Neptunes and gas giants are only observed in 5% of the recognized extrasolar planetary systems (Fressin et al., 2013). Extra-solar planets have been detected not only around Sun-like stars, but also around a variety of stellar objects ranging from high mass white dwarfs to low-mass, cool M dwarfs as well as in orbit around multiple stellar systems (Roell et al., 2012; Dressing and Charbonneau, 2013). Based on statistical considerations regarding the small or/and low mass planets and planet candidates detected so far, it is expected that rocky planets in the HZ of other stars are numerous.

In the Solar System, the planets orbit close to the ecliptic plane on close to circular orbits. If we compare this with what is now known about planetary systems in general, it appears that the Solar System is a rather exceptional phenomenon and is not a representative proto type for other planetary systems. Extra-solar planets at intermediate to large orbital distances have been found on circular and on eccentric orbits, with an averaged mean eccentricity of about 0.29 (Udry and Santos, 2007). Many of these planets display high spin-orbit misalignments; i.e., the angle between the stellar rotation axis and the planets orbital plane, and some of these planets apparently orbit their host star on retrograde orbits (Fabrycky and Winn, 2009). Such observations are indicative for significant dynamical interactions during the early history of those planetary systems.

Judging from the diversity of planetary systems which feature planets and stellar objects of various compositions, sizes and orbital properties, one might expect a great diversity of possible conditions for habitability and life formation compared to the conditions fulfilled in the Solar System. However, our current knowledge is at present far away from describing and understanding such a context within the framework of the observed extra-solar systems.

## 3. Formation of a habitable planet in the Solar System

This chapter presents the models and results that were obtained during the course of the Alliance regarding the formation of the terrestrial planets in the Solar System. This overview aims to highlight our understanding of those processes that led to the formation of a roughly Earth-sized habitable planet at 1 AU, its survival on a quasi-circular orbit and its accretion of a small, but important, fraction of the mass from water-rich planetesimals coming from outside the inner part of the Solar System. Furthermore, the first ~1 Gyr of post-accretion bombardment of the terrestrial planets by asteroids and comets and its importance for the formation of an Earth-like habitable environment will be evaluated.

### 3.1 Migration of giants and formation of rocky planets

After the formation of the first small planetesimals inside the disk of gas and solids that surround the newly-formed star, the planetesimals start to move and migrate inside the disk. Over time they can collide with each other and form so-called planetary embryos. Their motion due to interactions with the gas disk is termed migration (for a review see Kley and Nelson, 2012). In some systems these embryos can merge with each other and, thus, grow to form the cores of giant planets.

It is known that in the Solar System the giant planets formed long before the Earth. In fact, the giant planets accreted substantial amounts of hydrogen and helium and, therefore, they should have completed their formation before the disappearance of the gas in the proto-planetary disk. From observations of disks around stars of known age, the removal of the gas is inferred to occur within just a few Myr (Haish et al., 2001) after stellar formation. In contrast, radioactive chronometers indicate that the Earth took 30-100 Myr to form (see Kleine et al., 2009 for a review).

During their migration through the disk, the planetisimals and planets interact with each other. Obviously, the most massive planets strongly influence the final state of the system. Therefore, with regard to habitability it is important that the giant planets follow orbits which enable the presence of terrestrial-sized planets on stable orbits within the HZ. Our results (see 4.3) suggest that such a condition is fulfilled only rarely. However, the presence of giant planets could under certain circumstances favour the formation of Earth-like planets, because the giants could scatter asteroids from outside the snowline (i.e. the astronomical distance from the star beyond which $H_2O$ is stable as ice) into the inner Solar System and, thus, deliver volatile rich material onto objects within the HZ.

This consideration suggests that a good model of terrestrial planet formation in those planetary systems where giant planets are present should first address the origin and evolution of the giant planets, because they set up the "environment" in which the terrestrial planets are eventually formed. The focus will be on a dynamical evolution model, because a static orbital planetary system evolution does not conclusively explain all properties of the modern Solar System, such as the small mass of Mars or the chemical composition of Earth.

**3.1.1 Migration of the giant planets**

First we will discuss which processes caused the giant planets in the Solar System to migrate through the planetary disk, first inwards and then outwards. The scenario with a reversed direction of migration was nicknamed "Grand Tack" (Walsh et al., 2011). An important clue to address this question comes from the pioneering work of Masset and Snellgrove (2001). These authors performed the first simulation of the evolution of Jupiter and Saturn in a disk of gas. Jupiter is massive enough to open a deep gap in the disk and it settles into slow, type-II-migration, whereby the planet basically follows the viscous evolution of the disk (Lin and Papaloizou, 1986). The smaller Saturn is unable to fully empty its co-orbital region and therefore migrates inwards much faster (Masset and Papaloizou, 2003). Saturn therefore approaches Jupiter until the orbits of the two planets become locked in their 2:3 mean motion resonance. Both planets then occupy a common gap, which has important consequences for their further migration. The gravitational interaction with the inner disk exerts a torque on Jupiter and pushes it outwards, whereas the outer disk exerts a torque on Saturn and pushes it inwards. The torques are proportional to the square of the planetary masses, so that the outward directed torque from the inner disk felt by Jupiter is stronger than the inward directed torque from the outer disk felt by Saturn. The net result is a joint outward migration, where the stabilizing resonant relationship between their orbital periods is kept constant.

Morbidelli and Crida (2007) showed that, as already conjectured by Masset and Snellgrove (2001), the process of outward migration is possible only if the mass of the outer planet is smaller than that of the inner one (mass ratios between ¼ and ½ being "ideal", which nicely encompasses the real Saturn/Jupiter mass ratio of 1/3). To date, this is the only explanation for the current distant orbits of the giant planets in the Solar System. Alexander and Pascucci (2012) showed in simulations that the "pile-up" of giant planets around 1 AU can be explained due to photo-evaporation of the gas disk during the late stages. This mechanism does not explain the distant orbits of the giant planets in the Solar System. Thus, if the Grand Tack scenario holds true, it would suggest that the characteristic absence of giant planets in the inner Solar System is not a property of planetary systems in general. The theory further suggests that giant planets do not generally reverse their inward migration and, therefore, usually lead to systems having giant planets situated in or close to the HZ in systems with the following properties:

- only one giant planet or

- with multiple giant planets, but having a "reversed" mass ratio compared with the Solar System (i.e. the outer planet being the more massive) or

- with a too large temporal gap between the formation of the giant planets (precluding them from approaching each other sufficiently close to trigger outward motion)

Numerous research groups have applied independent hydrodynamic codes which have provided insight into the potentially wide range of possible giant planets migration scenarios during the time these giants are still embedded in the gaseous proto-planetary disk (see for instance Ward, 1997; Crida and Morbidelli, 2007; Pierens and Nelson, 2008; D'Angelo and Marzari, 2012). These works demonstrated that the previously held theory i.e., static perturbations of the giant planets from their current orbits while the Earth and its precursors formed, probably represents a rather simplified approach. In fact, numerical models of terrestrial planet formation, starting from a disk of planetesimals and planetary embryos extending from the Sun to the current orbit of Jupiter and assuming the giant planets on fixed orbits, consistently fail to reproduce one characteristic of the real terrestrial planet system: the small mass of Mars (Chambers, 2001; Raymond et al., 2009).

**3.1.2 Observational evidence for the Grand Tack**

Hansen (2009) convincingly showed that a key parameter relevant to the "quest" to obtain a small Mars is the radial distribution of the solid material in the disk: an outer edge of the disk consisting of planetary embryos and planetesimals is needed, placed at about 1 AU. This result motivated us (see e.g. Walsh et al., 2011) to combine terrestrial planet accretion with giant planet migration. Walsh et al. built their model on the inward-then-outward migration scenario – also known as "Grand Tack" - for Jupiter described above. They remarked that a reversal of Jupiter's migration at 1.5 AU would provide a reasonable explanation for the existence of the outer edge at 1 AU of the disk of solids. Consequently, the mass distribution of the terrestrial planets is statistically reproduced (see Fig. 2).

A crucial constraint of the Grand Tack scenario is the necessary survival of the asteroid belt. Given that Jupiter should have migrated through the asteroid belt region twice, first inwards, then outwards, one could expect a scattered asteroid belt that is either highly depleted in objects or completely empty. However, the numerical simulations by Walsh et al. (2011) show that the passage of the giant planets first dispersed the vast majority of objects in the asteroid belt, but then, as Jupiter leaves this region for the last time while moving back outwards, the asteroid belt gets re-populated by a small fraction of planetesimals that were scattered initially during the early inward motion of the giant planets. Thus, the Grand Tack scenario is consistent with the existence of the asteroid belt and, as discussed in Walsh et al. (2011), can explain its main orbital and physical properties.

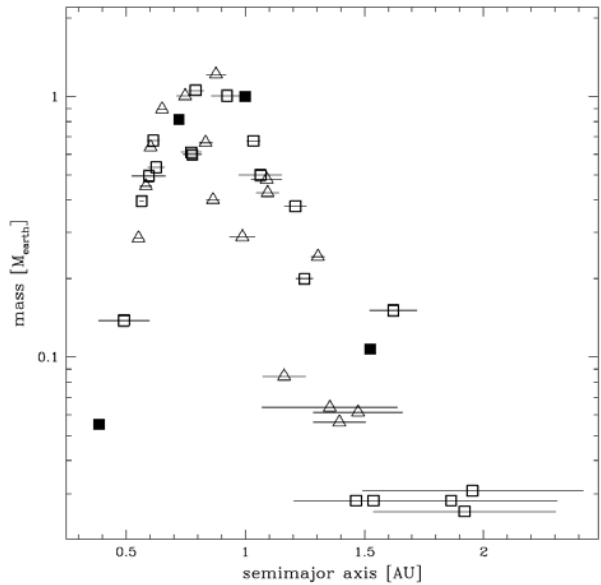

*Figure 2* The mass distribution of the synthetic terrestrial planets produced in the Walsh et al. (2011) simulations. The open symbols represent the planets produced in different runs starting from different initial conditions. The horizontal lines denote the perihelion-aphelion excursion of the planets on their eccentric final orbits. The black squares show the real planets of the Solar System. The large mass ratio between the Earth and Mars is statistically reproduced.

The outer two ice giants, Neptune and Uranus, are not massive enough to influence the motion of Jupiter and Saturn and their influence on the structure of the inner Solar System is minimal in the Grand Tack scenario.

**3.2 The post-accretion heavy bombardment of the terrestrial planets**

The predicted modern orbital configuration of the giant planets resulting from the Grant Tack scenario, however, differs from the current planetary system. The giant planets are predicted to form on a much more compact configuration. Therefore, some additional process is required in order to shift Saturn, Uranus and Neptune outwards to their current position. Such a massive rearrangement would result in the dispersal of the planetesimal disk which would lead to a spike in the impact rate of the terrestrial planets and the Moon. Key questions for the present study are therefore: when did this event occur?, can we use the lunar impact crater record to address this question? and, how can this event be understood in terms of processes in the planetary system? We therefore briefly review the impact record of the terrestrial planets.

**3.2.1 Sample record of the bombardment history**

Visible evidence of massive impact cratering events is retained on the ancient crust of Mercury, Moon and Mars. The flux of projectiles was investigated and quantified via crater density (i.e., the size-frequency distribution of craters per square kilometre) shaping the bombarded landscapes (e.g., Öpik, 1960; Baldwin, 1971). In a quasi-static system one assumes that the crater production function (related to the velocity and size-frequency distribution of the impacting projectiles) has not changed over time. This assumption however has been debated for some time; in particular the timing of a potential change in the crater production function remains unclear (Strom et al., 2005; Werner, 2008; Fassett et al., 2012; Marchi et al., 2012). In addition, it is difficult to assess the shape of the crater size-frequency distribution at such an early stage of the Moons history. If the impact flux is sufficiently high (after some time it will be), the surfaces become saturated, implying that the continuous flux of projectiles erases earlier crater records by subsequent cratering and the surface crater density reaches a steady-state (i.e., Gault, 1970; Richardson, 2009). Some groups of researchers found that the surface in the lunar highlands was saturated by craters larger than 2 to 20 km diameter (e.g. Hartmann 1984, and references therein), while other groups (Neukum and Ivanov, 1994, and references therein) consider that the crater size distribution follows the production function for such crater sizes. Despite this controversy, the bombardment of the Moon was clearly orders of magnitude higher during the first 1 – 1.5 Ga compared to the last 3 Ga (Neukum et al., 2001).

The impact flux record was calibrated to absolute ages from lunar samples obtained by the Apollo and Luna missions (e.g., Schaefer et al., 1976; Stöffler and Ryder, 2001). The stratigraphically oldest basins for which the assignment of absolute ages has been attempted by correlating with lunar sample material are the lunar front side basins Serenitatis, Nectaris, and Imbrium. However, there is lack of agreement within the lunar community regarding the lunar impact chronology. Two essentially different bombardment scenarios have been proposed. Large numbers of rocks collected from the equatorial near side of the Moon by the Apollo and Luna missions include impact-reset ages of roughly 3.9 Gyr. This observation suggests either 1) several basins formed around 3.9 Gyr (Turner et al., 1973; Tera et al., 1974; Stöffler and Ryder, 2001), or 2) all 3.9 Gyr impact ages are mainly related to the Imbrium impact event that delivered material to all Apollo landing sites (Baldwin, 1974; Tera et al., 1974; Haskin et al., 1998; Warren, 2003; Fernandes et al., 2013a). Scenario 2 is in agreement with a monotonically declining flux of impactors during the "Early Bombardment" and supports the proposed exponential decay rate (Neukum and Ivanov, 1994, Hartmann and Neukum, 2001). Scenario 1 is in favour of a pronounced spike of the impactor flux. Some works (e.g., Turner et al., 1973; Tera et al., 1974; Ryder, 1990) have argued that several or even all lunar basins formed during a brief time interval centred around 3.9 Gyr, i.e., ~600 Myr after terrestrial planet formation. This impact surge was referred to as a "Late Heavy Bombardment (LHB)" or "terminal lunar cataclysm". Intermediate views argue that the LHB was protracted over a few hundred million years (Turner, 1979) and started ~4.2 - 4.1 Gyr ago (Hartmann et al. 2000; Marchi et al., 2012; Morbidelli et al., 2012). To avoid further confusion we will refer to the time at which all lunar basins formed as the "Heavy Bombardment Eon" (HBE) encompassing the Pre-Nectarian and Nectarian periods and the Lower Imbrian Series of the selenologic time table. Considering that on the Moon the relatively high bombardment rate appears to tail off at 3.5 to 3.0 Ga ago (Neukum et al., 2001), the HBE therefore extends into the terrestrial Archaean. Whether such an Archaean intense bombardment was continuous (Bottke et al., 2012) or spiky (Fernandes et al., 2013a), and its relevance for the evolution of the environmental conditions can be addressed by investigating Archaean spherule layers which represent the oldest traces of impacts on Earth (Low et al., 2003; Goderis et al., 2013).

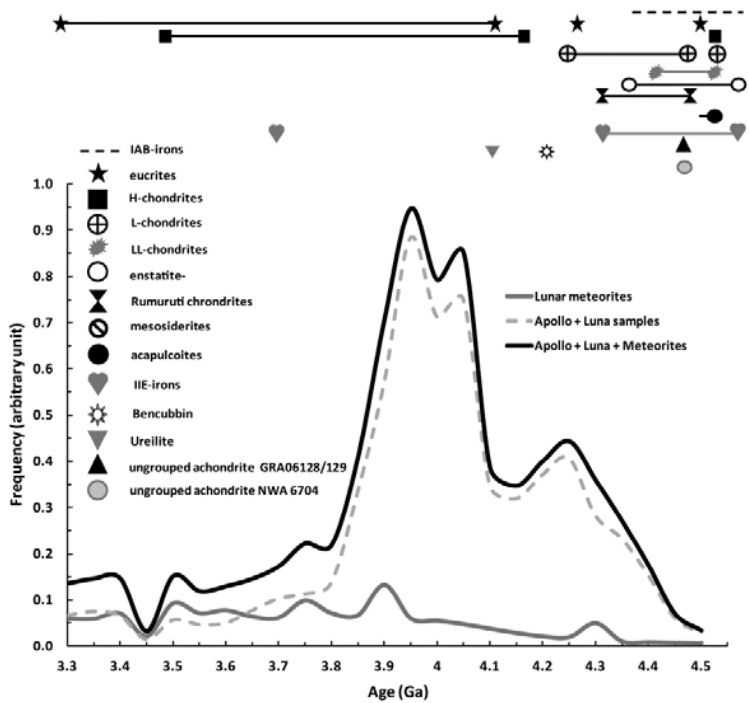

*Figure 3* Summary of impact ages of samples from Apollo and Luna missions and lunar meteorites (modified after Fernandes et al., 2013a) compared with impact ages obtained from other planetary material. The Gaussian probability curves were calculated by using the age and error of each sample and combined in bins of 0.05 Gyr (50 Myr), which is representative of the average error in $^{40}Ar/^{39}Ar$ age determination. The normalized Gaussian curve calculated for each age bin (age column) was obtained by taking into consideration the width of the Gaussian curve calculated and the measured uncertainties. Each column was added and the result normalized to an integer. This integer was used to normalize all curves shown in the diagram. Thus, the height of the curve/peak (y-axis) reflects the relative frequency of a certain age (x-axis). Where necessary, the age was corrected for monitor age and decay-constant. The thick black line is the cumulative impact ages for Apollo, Luna, and meteorites. The impact ages for the planetary ages: a) IAB iron-meteorites (Vogel and Renne, 2008); b) unbrecciated eucrites (Vesta?), Bogard and Garrison (2003); c) H-chondrites (Fernandes et al., 2006; Swindle et al., 2009 and references therein); d) L-chondrite (Turner et al., 1978; Bogard and Hirsch, 1980; Benedix et al., 2008; Weirich et al., 2011); e) LL-chondrites (Trieloff et al., 1989, 1994; Dixon et al., 2004); f) enstatite chondrite (McCoy et al., 1995; Dixon et al., 2003; Bogard et al., 2010); g) Rumuriti chondrite (Bogard, 2011 and references therein); h) mesosiderites at 3.9 Gyr (Bogard, 2011 and references therein); i) Acapulcoites (Bogard, 2011 and references therein); j) IIE iron silicates (Bogard, 2011 and references therein); k) Bencubbin meteorite (Marty et al., 2010); l) ureilitic meteorites (Bogard and Garrison, 1994); m) GRA 06129 (Shearer et al., 2010; Fernandes and Shearer, 2010); n) ungrouped achondrite NWA 6704 Fernandes et al., 2013b).

The controversial discussion regarding the lunar impact record derives from difficulties in relating individual samples collected by the Apollo and Luna missions from the impact gardened lunar surface to specific basins. This issue is most problematic for those basins older than 3.9 Ga because the degree of redistribution, crushing and mixing of crustal material strongly increases going back in time over the HBE. This issue is less contentious for dating basaltic flows and ejecta layers of relatively recent small lunar craters as compared to the ancient basin sized impact structures. The younger surfaces superimpose stratigraphically the older units and are less disturbed by impact gardening.

Nevertheless, using high resolution $^{40}$Ar-$^{39}$Ar step heating techniques together with a detailed petrological description of Apollo 16 and 17 samples together with literature data we (Fernandes et al. 2013a; and references therein) showed that the lunar impact record dates back to at least 4.3 Gyr. Some Apollo 17 samples and especially the Apollo 16 anorthositic rock samples collected along the North Ray crater ejecta show a high abundance of impact ages of ~4.2 Gyr (Fig. 3) that can reasonably be related to one or several lunar near side basins such as Tranquilitatis, Serentiatis or Nectaris. All of these basins delivered abundant material to the Apollo 16 landing site (Petro and Pieters, 2006, 2008) and are stratigraphically younger than the stratigraphically oldest named lunar basin (Wilhelms, 1987), the South Pole Aitkin. Consequently we (Fernandes et al. 2013a) argued that the outstanding large South Pole Aitken basin is ≥ 4.3 Gyr as given by the age of the oldest lunar impact melt rock identified in our study. In addition, it was shown that there is no compelling reason to maintain a 3.9 Gyr age for the Nectaris basin (Baldwin, 2006; Norman et al., 2010; Fischer-Gödde and Becker, 2012; Fernandes et al., 2013a). This removes the need to squeeze all basins that formed in the Nectarian period (and possibly pre-Nectarian) into a 100 Myr short time interval around 3.9 Gyr ago. A rather low impact flux onto the Earth-Moon system around 3.9 - 3.7 Gyr ago is essentially consistent with the low concentrations of extra-terrestrial material (Ir concentrations and Cr isotopic ratios) reported from ~3.8 Gyr old meta-sedimentary sections from Greenland (Koeberl et al., 2000; Frei and Rosing, 2005).

The implication of an intense on-going bombardment through the entire HBE is essentially consistent with the impact record deduced from meteorites originating from the asteroid belt. Some meteorites like the howardite, eucrite and diogenite (HED) achondritic meteorites (thought to have originated from the crust of the differentiated asteroid Vesta) and H chondrites (high iron ordinary chondrites; i.e., reduced) show a dominance of impact ages between 4.1 and 3.5 Gyr ago. Several other meteorite types from the asteroid belt including the enstatite chondrites (Mg-rich pyroxenes and metal, e.g., highly reduced) and the LL chondrites (low iron low metal ordinary chondrite, e.g., relatively oxidized) show a range of impact ages older than 4.2 Gyr but none in the time between 4.1 and 3.5 Gyr ago (Fig. 3).

The collision history of the Solar System appears therefore to be quite complex, and possibly varies between different regions of the asteroid belt. This is suggestive of the past existence of other projectile sources (planetary left-overs, asteroidal belt(s), debris disks etc.) that no longer exist, but which provided the source of material that impacted the inner terrestrial planets. These sources could include very massive heliocentric and maybe mostly achondritic (high-siderophile element poor, and relatively oxidized planetary mantel material) debris disks from planet sized collisions, with the latest being a putative giant impact that formed the Earth-Moon system (Schlichting et al., 2012; Jackson and Wyatt, 2012). Based on the overall impact record in lunar rocks and meteorites from the asteroid belt we (Fernandes et al. 2013a) proposed that the impact flux during the entire history of the Earth-Moon system is characterized by several episodes of increased bombardment. The intensity and temporal duration of these spikes likely varies for the different events that triggered the increased delivery of projectiles to Earth-crossing orbits.

Samples with ages that can be directly linked to a specific pre-Nectarian basins would be required to quantify the overall time evolution of the complex bombardment record. The 14 or 15 basins from the Nectarian period and the Imbrium series (Wilhelms, 1987; Spudis et al., 2011) formed between ~4.2 and ~3.7 Gyr ago. In summary, the strong arguments against the classical lunar sample chronology with several basins forming around 3.9 Gyr ago impose severe boundary conditions for the "Grand Track" scenario discussed in Section 3.1. The key question is how to shift the orbital configurations of the gas giants in the Solar System out to where they are without producing a narrow and very intense bombardment of the terrestrial planets.

**3.2.2 The "Nice model"**

Recall that, after the inward and then outward migration of Jupiter, described in the Grand Tack model (Section 3.1), the four giant planets in the Solar System remained on quasi-circular, resonant, compact orbits (from 5.5 AU to ~17 AU, which is a much more compact range than their present-day values. Further out (at around 35 AU) a disk of planetesimals remained, containing 30-50 Earth masses. The question is: how did the giant planets evolve from this post-Grand Tack configuration to their current orbital configuration? Addressing this question is the purpose of the "Nice model"

It is well known that resonances have a strong stabilizing effect on the dynamics. However, due to the gravitational interactions with the disk of planetesimals, the planets were eventually extracted from their multiple resonances (possibly after several 100's Myr). The extraction is caused by scattering planetesimals, which gives kicks to the planet that cause random walks in the libration amplitude. If the latter becomes too large, the planets escape from their resonance. Once the planets are extracted from their mutual resonance its stabilizing effect ends. The planets became rapidly unstable, because they were relatively close to each other. The planetary orbits became chaotic and started to approach each other. At some point, both Uranus and Neptune were scattered outward, onto large eccentric orbits (e = ~0.3–0.4) that penetrated deeply into the planetesimal disk, and dispersed it. Jupiter, on the other hand, moved slightly inwards to its current orbit. This inward movement of Jupiter happened on a very short timescale (10kyr – 100kyr), and, therefore, this scenario is called "jumping Jupiter" (Morbidelli et al. 2010), a nomenclature already introduced by Marzari and Weidenschilling (2002) for the origin of the eccentric orbits of giant extrasolar planets. The interactions with the planetesimals damped the planetary eccentricities of Uranus and Neptune, stabilizing the planetary system once again, and forcing a residual short radial migration of the planets. By the time that most of the planetesimal disk was eliminated, the giant planets had already acquired their current orbits (Tsiganis et al., 2005; Nesvorny and Morbidelli, 2012)

The "Nice model" can potentially explain the origin of a Late Heavy Bombardment (LHB) of the inner Solar System, whereby the dynamical instability of the giant planets and the consequent dispersal of the planetesimal disk caused a raised impact rate onto the terrestrial planets and the Moon. Two key issues are: how big was this spike and when did it occur? In order to characterize the flux of projectiles onto the Moon efforts have been focused to constrain the actual evolution of the giant planets during their unstable phase.

In the Nice model, the projectiles bombarding the Moon are derived from two possible reservoirs: a disk made of comet-like objects that existed originally beyond the giant planets, and the asteroid belt. There is no obvious trace of a cometary bombardment on the Moon (Kring and Cohen, 2002; Tagle, 2005; Strom et al., 2005), which seems to imply that physical disintegration, possibly due to explosive ice sublimation (outbursts) or the quite common cometary splits (Boehnhardt, 2005), decimated the cometary population as it penetrated into the inner Solar System (e.g., Sekanina, 1984). Concerning asteroids, Morbidelli et al. (2010) showed that the jumping-Jupiter evolution only removed about 50% of the objects within the current boundaries of the main belt. This would have produced only 2-3 basins on the Moon, not enough to match the heavy bombardment constraints.

Therefore, additional sources of lunar projectiles are needed. This led to the development of the E-belt concept (Bottke et al., 2012) which stems from the realization that the current inner boundary of the asteroid belt (~2.1 AU) is set by the $\nu_6$ resonance, whose existence is specifically related to the current orbits of Jupiter and Saturn. More specifically, this resonance moves towards the Sun as the orbital distance between Jupiter and Saturn increases. Moreover, its strength depends on the eccentricities of Jupiter and Saturn. Before the giant planets changed their orbital configuration, Jupiter and Saturn were closer to one another and were on more circular orbits. Therefore, at that time, the v6 resonance was located beyond the asteroid belt and was much weaker. Hence, the asteroid belt could extend down to the actual stability boundary set by the presence of Mars; i.e., down to 1.7-1.8 AU, depending of the original eccentricity of the planet. Note that the existence of this putative extended belt population (E-belt) between 1.7-2.1 AU has not yet been demonstrated to be consistent with the Grand Tack scenario. The E-belt would have been almost fully depleted when the orbit of Jupiter, and the v6 resonance, "jumped" to their current location. According to our simulations of this process, until today only ~0.15% of the bodies from the E-belt would survive, making up the population of the Hungaria asteroids (a group of small and highly-inclined bodies at 1.8-2.0 AU). This allows for a rough calibration of the original E-belt population. From this, it has been concluded that the E-belt population could have caused an additional 9-10 impact basins on the Moon, over a period of about 400 Myr. Therefore, given that the last basin on the Moon formed ~3.75 Gyr ago, the Nice/E-belt model argues that the first of the LHB basins formed 4.1-4.2 Gyr ago. One of these basins can be identified to be Nectaris, given that it was the ~14$^{th}$ or 15$^{th}$ from last to form (Wilhelms, 1987; Spudis et al., 2011). Moreover, Marchi et al. (2012) found arguments that the projectiles impacting the Nectaris basin derived surfaces (basin floor and ejecta blanket) had on average higher velocity than those that cratered pre-Nectarian terrains such as the feldspatic highlands or the South Pole Aitken terrain. This is consistent with Nectaris being formed within the bombardment caused by projectiles from the E-belt, because these projectiles are characterized by on average higher impact velocities compared to those from the previous period.

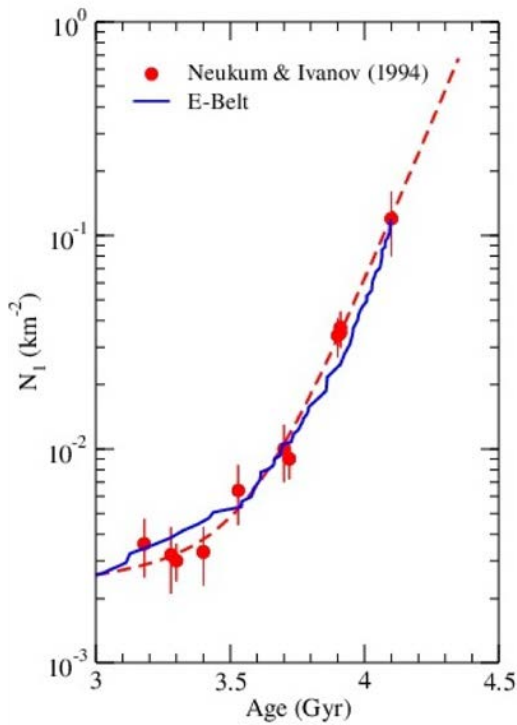

*Figure 4* *The dots vertical error bars show the number of craters larger than 1 km in diameter per square kilometre of lunar surface as a function of surface age, according to Neukum and Ivanov (1994). The distribution of dots can be interpolated by the function $N_1$= 5.44 $10^{-14}$ exp (6.93 t-1)+8.38 $10^{-4}$ t, where t is the age of the surface in Gyr (dashed curve). Beyond 3.6 Gyr, the fit is dominated by its exponential part. The last terrain of "known age" (the rightmost dot) is Nectaris, with an age of 4.1 Gyr. Beyond this age, the fit is extrapolated towards the lunar formation age, ~4.55 Gyr ago. The blue curve shows the decay of the bombardment rate in the Nice/E-belt model.*

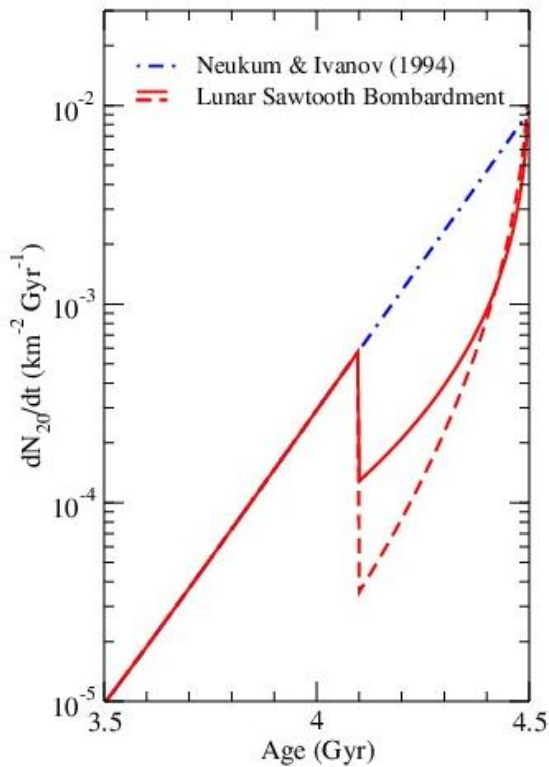

*Figure 5 The lunar bombardment rate as a function of time (measured in number of craters larger than 20 km produced per $km^2$ and per Gyr). The blue dash-dotted line is the exponential decay assumed by Neukum and Ivanov (1994). The red curve results from combining a post-accretion bombardment of leftover planetesimals in the 4.1-4.5 Gyr time range and the asteroid main belt and E-belt bombardments for t <4.1Gyr. The solid and dashed curves for t >4.1 Gyr correspond to the two analytic functions that bracket the early decay of the cratering rate, as discussed in Morbidelli et al. (2012). The graph assumes a 4.1 Gyr age for Nectaris to allow comparing the two calculated bombardment time lines. Not that the age for Nectaris may reasonably be shifted to 4.2 Gyr as discussed in Section 3.2 and shown in Figure 4.*

Morbidelli et al. (2012) showed that the declining bombardment with projectiles from the asteroid belt and the E-belt during the 4.1-3.5 Gyr period matches remarkably well the bombardment timeline envisioned by Neukum and Ivanov (1994) on the basis of the crater record in lunar geologic units with ages determined by lunar mission samples (Fig. 4). To model the lunar bombardment before 4.1-4.2 Gyr ago, Morbidelli et al. (2012) studied the declining planetesimal population leftover from the terrestrial planet accretion processes in the Grand Tack scenario. These authors calibrated the total projectile mass accreted by the Moon using constraints from the highly-siderophile element (HSE) content in the lunar crust and mantle. They concluded that the Moon's bombardment had a saw-tooth profile, with a single moderate uptick at 4.1-4.2 Gyr ago (see Fig. 5). This stands in sharp contrast to the prominent impact spike usually shown in sketches of the lunar cataclysm, instead it is in broad agreement with the scenario of "weak cataclysm" promoted in Turner et al. (1979) or in Fig. 3 of Hartmann et al. (2000). During the Alliance two groups (Morbidelli et al., 2012; Fernandes et al., 2013a) independently concluded that: 1) the outstanding large lunar basin South Pole Aitkin is older than 4.3 Ga and part of another projectile population compared to those projectiles that formed some of the younger lunar impact basins; 2) there was no extremely intense and brief (100-200 Myr duration) spike in the impact flux onto the Earth-Moon system at ~3.9 Ga ago (terminal lunar cataclysm scenario); and that 3) the impact flux during the Heavy Bombardment Eon (HBE) is more complex than a simple exponential decay.

## 4 Discussion

### 4.1 Water on Earth

Of crucial importance for a habitable Earth is the delivery of water and organic material. Laboratory analyses show a clear correlation between the water content in meteorites and the heliocentric distance of their parent bodies (see Fig. 6). Although there is the theoretical possibility that water-rich planetesimals formed in the hot regions of the disk by water-vapour absorption on silicate grains (Muralidharan et al., 2008; King et al., 2010), the empirical evidence for such a correlation strongly suggests that the planetesimals in the terrestrial planet region were extremely dry, thus raising the question of how the water came to the Earth.

According to the Grand Tack scenario (Section 3.1) the giant planets implant primitive, water containing planetesimals of carbonaceous chondritic like composition from the outer and cooler regions into the asteroid belt. In the Walsh et al. (2011) simulations, for every primitive planetesimal implanted in the outer asteroid belt, 10–30 planetesimals ended up on orbits that cross the terrestrial planet forming region, with a total mass of $3–11 \times 10^{-2}$ Earth masses. However, the accretion of water in this scenario is not uniform throughout the process of planet accretion but it accelerates during the end of this phase.

O'Brien et al. (2010) showed that, in this situation, the Earth could have accreted about 2% of its mass from these objects, and, thus, enough to supply a few times the current amount of water on/in the Earth. It is assumed that the primitive planetesimals consist of 5-10% water by mass and according to estimates the water content of Earth ranges from $2.5 \times 10^{-4}$ to $2 \times 10^{-3}$ Earth masses (Lecuyer, 1998; Marty, 2012). These water rich bodies would be of the same type as those captured in the asteroid belt as C-type asteroids. Since C-type asteroids are the parent bodies of carbonaceous chondrites and the D/H ratio of their water is on average like that of the Earth, this model also explains the isotopic composition of Earth's water. It is purely semantic to call this accretion "wet", since most of water present on Earth today was delivered during accretion, or to call it "dry", since it was a small amount compared to carbonaceous and ordinary chondrites (see Fig. 6).

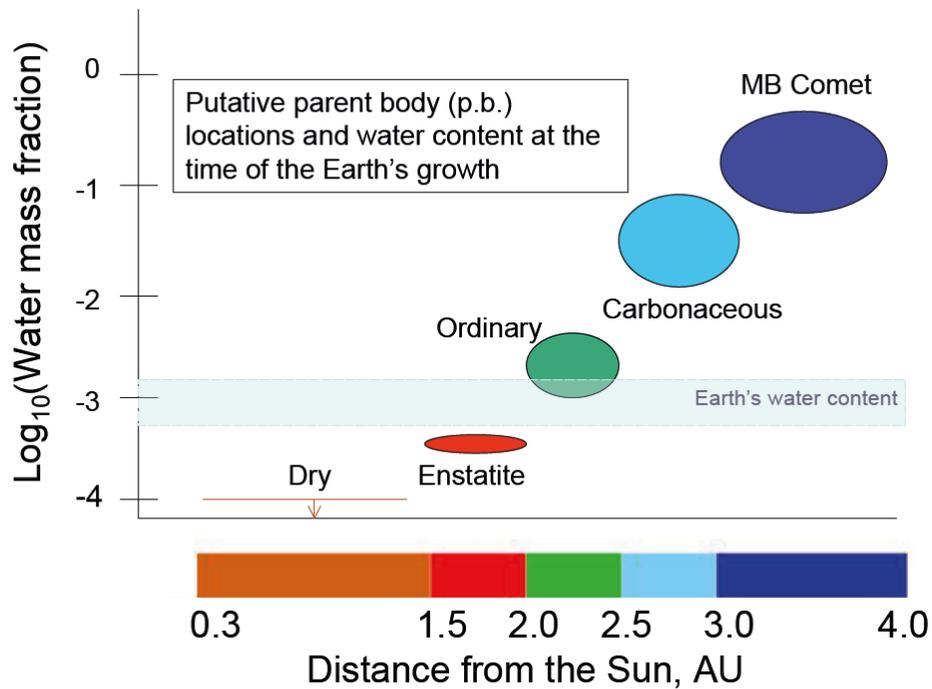

*Figure 6* CI and CM meteorites are the most water rich ones with water amounts of about 5 to 10% of their total mass (Robert and Epstein, 1982; Kerridge, 1985). They are expected to come from C-type asteroids, predominantly in the asteroid belt and possibly accreted even further out (Walsh et al., 2011). Water in ordinary chondrites amounts for only 0.1% of the total weight (Robert et al. 1977; Robert et al., 1979; McNaughton et al., 1981), or a few times as much (Jarosewich, 1966); they are spectroscopically linked to S-type asteroids, predominant between 2 and 2.5 AU. Finally, enstatite chondrites are very dry, with only 0.01% of their total mass in water; they are expected to come from E-type asteroids, which dominate the Hungaria region in the very inner asteroid belt at 1.8 AU.

Other authors argue that significant masses of water and other volatiles were mainly delivered after the main accretion phase via impacts as a "late veneer" (Albarede, 2009). According to the results described in Section 3.2 the heavy bombardment of the fully developed Earth-Moon system was characterized by two episodes: an early (>4.3 Gyr) one due to the decay of the planetesimal population left over from the accretion of terrestrial planets and a late (<4.2 Gyr) one described in the "jumping" Jupiter scenario (Section 3.2). In the previously described model projectiles that formed in the inner Solar System, such as dry enstatites or comparatively wet ordinary chondrites dominated both episodes of the Heavy Bombardment Eon. Comets are only marginally represented in this impactor population that reached the habitable zone (HZ) and, thus, can be neglected. This is in agreement with geochemical findings on the Moon (Kring and Cohen, 2002; Tagle, 2005).

Given the magnitude of the bombardment and the physical nature of the projectiles, it can be concluded that the heavy bombardment onto the Earth-Moon system did not favour the emergence of habitable conditions by delivering the fundamental ingredients for life e.g., water and organics. These ingredients were more likely delivered previously, during the accretion phase of the planet. This scenario appears consistent with the chemical and isotopic composition of Earth (Halliday, 2013). The late (<4.2 Gyr) episode of the heavy bombardment was not massive enough to sterilize the planet because the bombardment time line described here includes less basins and is spread over a longer time interval compared to the bombardment time line on which Abramov and Mojtzsis (2009) excluded sterilization. Because the population of planetisimal left over from the planet accretion phase, including the largest projectiles, mainly arrived before 4.4 - 4.3 Gyr it is feasible that the heavy bombardment became weak enough to allow for a habitable Earth already at 4.4 - 4.3 Gyr ago.

## 4.2 Impacts and atmospheres

Even though the components of the HBE (Section 3.2) may not have delivered considerable amounts of organics and water to the terrestrial planets these impacts modified their atmospheres. Clearly atmospheric evolution, hence composition and density is expected to strongly influence the habitability and bio-signatures (Grenfell et al., 2013).

To evaluate the consequences of impacts on composition and density of atmospheres new hydrocode simulations with different initial conditions were performed in the framework of the Alliance. We (Shuvalov et al., This issue) studied dense steam atmospheres proposed for an atmosphere around the early Earth (Zahnle et al., 2007). It was found that atmospheric erosion may dominate over retention for high impact velocities (>30 km/s) and an atmospheric pressure of 200 bar. In this case asteroids and comets a few kilometres in diameter disrupt in the atmosphere and the aerial bursts result in the escape of air masses comparable to the mass of the impacting projectile.

For low density (<200 bar) atmospheres the effect of impacts during the late part of the heavy bombardment was studied by us (de Niem et al. (2012) using the impactor fluxes predicted by the Nice model (Morbidelli et al. 2012), the addition of volatiles from the impactors dominates over loss processes. Thus, these impacts result in atmospheric growth for Earth and Mars even in those cases where the volatile content of the impactor is less than 1%.This leads to atmospheres of a few bars. In fact, whether the bombardment is rather short and intense (as in the classical LHB assumption of a distinct peak), or spiky and spread over a longer period of time (see Section 3.2) does not significantly affect the calculated consequences of impacts on the density of Earth's atmosphere. The impact simulations however showed that a few very large impact events (impactor sizes of a few hundred km diameter) on planetary surfaces dominate the evolution of atmospheres over numerous smaller impact events.

The results presented here suggest that the delivery of volatiles by the flux of impactors generally leads to a more or less efficient growth of atmospheres. With increasing atmospheric thickness the net gain decreases and may finally result in the loss of mass once a certain atmospheric density is exceeded. This can be considered as some sort of self-regulation process. Nevertheless, based on numerical models on impact cratering it appears that neither a giant impact (Genda and Abe, 2003; Melosh, 2003) nor an intense bombardment (de Niem et al., 2012; Shuvalov et al., This Issue) efficiently removes the volatiles from a „wet" accreted Earth. Other processes such as magmatic outgassing and atmospheric erosion by enhanced soft X-ray and EUV flux of the Early Sun (Erkaev et al., This Issue) that can affect the composition and density of the atmosphere around planetary embryos or planets are not a subject to this paper.

**4.3 Habitability of planets as a result of specific formation scenarios**

By extrapolating from the understanding regarding the formation of terrestrial planets in the Solar System described in the previous sections it is possible to speculate on the abundance of Earth like worlds in extra-planetary systems, or more generally on the possible structures of planet systems hosting terrestrial planets.

As mentioned in Section 3.1, a key but non-generic characteristic of the Solar System is the presence of giant planets that are all at large heliocentric distances and presumably never penetrated closer than 1.5 AU. Another important characteristic are the quasi-circular orbits of the giant planets in the Solar System , showing eccentricities unlike those of most extra-solar giant planets known to date (see Section 2.3). Below are listed four general case studies, in which the dynamics and evolution of the system significantly differ, compared to the Solar System.

**Case studies**

*Case 1) Giant planets that never migrated through the habitable zone (HZ) – the Solar System example*

In planetary systems with the giant planet(s) that remained on quasi-circular orbit(s) in the outer part of the system, the accretion in the inner part of the disk proceeds along a well-studied path. Planetesimals accrete to more massive objects, known as planetary embryos, through the subsequent processes of runaway (Greenberg et al., 1978) and oligarchic (Kokubo and Ida, 1998) growth. Depending on the initial surface density of solid material in the disk these planetary embryos acquire masses similar to the masses of Mars or the Moon. Once the gas in the disk disappears, the system of embryos develops dynamical instabilities forcing the embryos to acquire more eccentric orbits that cross each other and, thus, allowing for collisional growth on timescales of several millions of years (Chambers, 2001). This process of mutual giant collisions eventually leads to a few planets with masses comparable to that of the Earth (Raymond et al., 2009). The delivery of water-rich material from the vicinity of the giant planets seems to be a quite generic process (Morbidelli et al., 2000; Raymond et al., 2004; 2006; O'Brien et al., 2006). The exact migration behaviour of the giant planets in the outer system is not expected to play a crucial role, as long as their orbital eccentricities remain small. From the viewpoint of the accretion of Earth, but not from the viewpoint of Mars!, the growth history of the terrestrial planets in previous models assuming giant planets on fixed orbits is qualitatively similar to those in the *Grand Tack* scenario. The HZ is narrow (~1 – 1.7 AU; Kopparapu et al., 2013) and, therefore, not all planetary systems will develop an Earth-like habitable planet. However, it can be expected that among the entire diversity of planetary systems that may form in this way Earth-like habitable planets are not rare.

*Case 2) Giant planets that migrated into/through the habitable region*

The migration of giant planets into the HZ causes disruption of the planetesimal disk. This disruption occurs via two mechanisms. One mechanism is that many planetesimals get captured in mean motion resonances with the migrating planets and are forced to migrate together with the resonances. This mechanism can drag a substantial fraction of the disk material out of the HZ, towards the star. Those planetesimals captured in these inward moving resonances can still accrete with each other to form terrestrial planets, but they would inevitably be too close to the star to be habitable (Fogg and Nelson, 2005; Raymond et al., 2006). The second mechanism is that those planetesimals not captured in the resonance are eventually scattered and dispersed by the giant planets. Those giant planets that remain in/near the HZ stir and deplete their neighbourhood and, thus, prevent the accretion of terrestrial planets in the same zone. In a similar way Jupiter stirred and depleted the asteroid belt, preventing the accretion of another planet in the belt. However, even in this scenario the formation of Earth-like habitable worlds is in principle possible. First, the moons of giant planets inside the HZ could be habitable. Second, the giant planet may migrate fast through the HZ, and leave it while continuing its approach to the star. In this scenario it is possible that enough of those planetesimals scattered behind the orbit of the giant planet, have their orbits re-circularized by the drag of the gas still present in the disk and eventually generate terrestrial planets after the gas is dispersed by the proto-star (Raymond et al., 2006) in a process similar to that discussed above for case (1).

*Case 3) Giant planets that developed large orbital eccentricities*

Most extrasolar giant planets detected so far are close to their host star and/or have highly eccentric orbits (see Section 2.3). It is believed that these eccentricities developed in the system during a phase of orbital instability and mutual scattering of the giant planets (Moorhead and Adams, 2005; Juric and Tremaine, 2008; Beaugé and Nesvorny, 2012). The increase of eccentricity to very large values for the most massive planets has devastating effects for the rest of the system. Through a mechanism of secular perturbations, the eccentricities of all objects are forced to undergo large oscillations. This is true independently from the value of the giant planets semi-major axes, because secular perturbations are effective even at long distances. Simulations by Raymond et al. (2011) showed that a system of terrestrial planets that are growing as described for case (1) become immediately destabilized when the giant planets jump onto eccentric orbits. Either all terrestrial planets are removed, because they acquire very eccentric orbits and collide with the central star or they intersect the trajectories of the giant planets which then eject the terrestrial planets onto hyperbolic orbits or only one terrestrial planet survives on a very eccentric orbit that may oppose habitability.

*Case 4) Planetary systems without giant planets*

Only about 5-10% of Sun-like stars, observed so far, harbour Jupiter-mass planets within a few AU (Vigan et al., 2012). Thus, one might think that the remaining ~90% of the Sun-like stars offer conditions favourable for the formation of terrestrial planets, including Earth-like habitable ones. In fact, it has been demonstrated that, in the absence of giant planets, the formation of terrestrial planets on orbits with moderate eccentricities is a generic process around Sun-like stars (Levison and Agnor, 2003). However, this situation may not be as simple in a system without massive planet(s). Whereas the formation of Jupiter-mass planets appears to be relatively rare, the formation of massive objects such as Uranus and Neptune seems to be more generic. The overall analysis of all candidates observed with the High Accuracy Radial velocity Planet Searcher (HARPS) suggests that ~30% of the extra-solar planets with mass smaller than 30 Earth masses orbit with periods shorter than 100 days around Sun-like stars (Lovis et al., 2009). From here on these objects will be referred to as "giant cores", because they might form the cores of the giant planets, that later grow by accreting gas, which is in contrast to Super-Earths that are mostly of rocky nature. When these giant "cores" are still embedded in a disk of gas, they are expected to reside near a "no-migration radius" that results from a balance between the various torques that they suffer from the disk (Lyra et al., 2010; Bitsch and Kley, 2011). However, as the disk evolves and the amount of gas is reduced, the no-migration radius moves towards the star; thus, the giant cores eventually migrate inward until the gas is substantially removed (Lyra et al., 2010). This process seems to be generic. In the Solar System the only apparent reason that inhibited the inward migration of Uranus and Neptune down to roughly 1 AU is that they have been retained in resonance with Jupiter and Saturn. In the Grand Tack scenario the planets Uranus and Neptune migrated outwards during the last phase of the lifetime of the disk. In case this understanding of the Solar System history is correct, then in most systems without Jovian-mass planets the HZ is eventually "invaded" by giant cores.

The effect of such an invasion regarding the forming terrestrial planets has not been studied in detail yet. We expect that most of the solid material originally in the HZ would be captured in resonances with the migrating giant cores and transported towards the star. Rocky planets can form from the material shepherd in resonances with the giant cores, as already shown by Fogg and Nelson (2005) for migrating giant planets, but the rocky planets would eventually be too close to the star to be habitable. The first evidence for such a process may be the Kepler-36 system, made of a super Earth just inside the orbit of a hot Neptune. These two planets have a large density ratio, because the super Earth is rocky whereas the hot Neptune should be made of rock and ice with a substantial atmosphere of light gases like that of Neptune (Carter et al., 2012). This implies that the invasion of giant cores into the habitable zone is "bad" for the formation of an Earth-like planet. We note that all the "terrestrial planets" formed in the celebrated planetary synthesis models (Alibert et al., 2004; Ida and Lin, 2008; Mordasini et al., 2009) are actually low mass "giant cores". That means these planets formed rapidly in a disk of gas and planetesimals, rather than planets such as the Earth that formed on a timescale of several tens of Myr, from mutual giant impacts among low-mass planetary embryos.

*Summary of the four cases:*

How often do cases (1), (2), (3) and (4) occur in planetary systems? Based on the current knowledge of extrasolar systems with giant planets, cases (2) and (3) seem to be more common; e.g., not favourable for Earth size planets in the HZ, whereas Solar System like case (1) seems to be the exception. However, due to observational biases, it is problematic to detect those giant planets hosting systems that are most similar to the Solar System. From the observational viewpoint, the census of extra-solar planetary system is still too limited. From the modeling point of view, a good understanding of the process of formation of the giant cores and Jovian-planets is required. The formation of giant planets currently is one of the major open problems in planetary science and will be further investigated during the Alliance. The key issue is to determine those locations within the disk in which the cores of the giant planet can form. Initial clues to this question can be extracted from migration maps of planets in disks compiled from simulations that consider realistic thermodynamics including viscous heating, radiative cooling and stellar irradiation (Bitsch et al., 2013).

## 5. Summary and conclusions

In many respects the system aspect is of significant importance for the formation of Earth-like habitable planets. This paper focused on dynamical processes in the Solar System related to the formation of habitable and rocky planets. Hydrodynamic processes shape the evolution of the molecular cloud that collapses to a solar nebula and finally to a central star with revolving planetary bodies (Tornow et al., This issue). The Sun was formed in a clustered environment that seems to be favourable for the formation of Earth like habitable planets (Section 2.2). Later, the mass, location and composition of rocky planets and their potential to host Earth-like habitable environments are strongly affected by the migration of the giant members of the planetary family.

The early evolution of the Solar System is described by the Grand Tack scenario (Section 3.1) that may be considered to be quite specific but is capable to generate a habitable Earth. In the Grand Tack scenario the two giant planets, Jupiter and Saturn, had an appropriate mass ratio that allowed them to migrated first inwards and then outwards through the inner part of the planetary system. In this particular case the giants migrated in such a way that they did not approached the HZ too close and maintained low to moderate eccentricities. Furthermore, during their outward migration, the giant planets triggered the delivery of primitive volatile-rich planetesimals from outer reservoirs into the asteroid belt and onto the growing Earth. Simulations showed that this process could explain the current water content of the Earth. However, in the "Grand Tack" scenario the delivery of water onto Earth occurred not uniformly throughout planet accretion, instead it accelerates towards the later stages.

The role of the subsequent evolution of the planetary system that left its fingerprint by means of the crater record was studied and described in the Nice model (Section 3.2). Assuming that the described single spiked saw-tooth like bombardment timeline derived from this simplified model is correct; it allows for some conclusions regarding the formation of the habitable Earth and the abundance of Earth like worlds in other planetary systems.

1) The Heavy Bombardment Eon (HBE) of the fully developed Earth-Moon system would be characterized by two episodes: an early one, due to the decay of the planetesimal population left over from the accretion of terrestrial planets and a late one, dominated by the E-belt projectiles. The early (first ~300 Myr) bombardment could be a generic feature of terrestrial planets that formed in any planetary system; in contrast, the late (after the initial ~300 Myr) bombardment is specific to the evolution of the giant planet in the Solar System, and, therefore, would be a non-generic feature. Both components of the HBE were dominated by projectiles that formed in the inner Solar System. These projectiles may have been either dry, such as enstatite or modestly wet such as ordinary chondrites. Given the magnitude of the bombardment and the physical nature of the projectiles, it can be concluded that ingredients necessary for life (e.g., water, organics) on Earth were probably delivered during the accretion phase of the planet, and not later during the HBE.

2) Even though the HBE did not deliver considerable amounts of organics and water to the terrestrial planets it could however have significantly modified their atmospheres (Section 4.2). Consequently, this bombardment that is a result of system inherent processes (Section 3.2) may in this way have shaped the habitability of terrestrial planets.

In conclusion, if the proposed reconstruction of the early evolution of the Solar System and the formation of terrestrial planets is correct, the evolution of giant planets supported the existence of a habitable Earth:

(1) by a formation and migration behaviour that eventually scattered water-rich planetesimals into the inner Solar System, including a part that could be accreted by the Earth;

(1) by their specific arrangement to avoiding migration too close to the HZ, but remaining tacked at about 1.5 AU. If they had reached 1 AU, the Earth would have probably been formed as small as Mars, which presumably would have precluded its long-term habitability;

(2) by remaining on quasi-circular orbits through their migration history. If they had become unstable and eccentric, they would have excited also the eccentricities of the orbits of the terrestrial planets, with dramatic consequences for the stabilities of the climatic conditions on those terrestrial planets.

Besides that there may be other scenarios to form an Earth-like habitable world the current evidences regarding the Earth advocate that the here described process operated at least once. From current knowledge on extra-solar planetary systems the most commonly observed structures of giant planet hosting planetary systems derived from those two cases with giant planets that either migrated into/through the habitable region or developed large orbital eccentricities (see discussion in Section. 4.3). But both cases may not lead to friendly conditions for habitable planets.

**Acknowledgements:** This research has been supported by the Helmholtz Association through the research alliance "Planetary Evolution and Life". V.A. Fernandes thanks the International Space Science Institute (ISSI) for hosting the team on updating the lunar chronology and for financial support by the Deutsche Forschungsgemeinschaft (DFG) FE1211/1-2. This research has made use of NASA's Astrophysics Data System. We also thank H. Lammer, the editor and an anonymous reviewer for their comments which helped us to improve previous versions of this contribution.